\documentclass[pra,showpacs, twocolumn, superscriptaddress]{revtex4}
\usepackage{amsmath}
\usepackage{amsfonts}
\usepackage{graphicx}
\usepackage{longtable}
\newcommand{\be}{\begin{equation}}
\newcommand{\ee}{\end{equation}}

\newcommand{\ba}[1]{\left(\begin{array}{#1}}
\newcommand{\ea}{\end{array}\right)}
\begin{document}

\title{Spin squeezing in symmetric multiqubit states with two distinct Majorana spinors} 
\author{K. S. Akhilesh }
\affiliation{Department of Studies in Physics, University of Mysore, Manasagangotri, Mysuru-570006, Karnataka, India}
\author{B. G. Divyamani} 
\affiliation{Tunga Mahavidyalaya, Thirthahalli-577432, Karnataka, India}
\author{Sudha} 
\email{arss@rediffmail.com} 
\affiliation{Department of Physics, Kuvempu University, 
Shankaraghatta-577 451, Karnataka, India}
\affiliation{Inspire Institute Inc., Alexandria, Virginia, 22303, USA.}
\author{A. R. Usha Devi}
\affiliation{Department of Physics, Bangalore University, 
Bangalore-560 056, India} 
\affiliation{Inspire Institute Inc., Alexandria, Virginia, 22303, USA.}
\author{K. S. Mallesh} 
\affiliation{Department of Studies in Physics,  University of Mysore, Manasagangotri, Mysuru-570006, Karnataka, India}
\date{\today}

\begin{abstract}
Majorana geometric representation of pure $N$-qubit states obeying exchange symmetry is employed to explore spin squeezing properties in the family  $\{{\cal D}_{N-k,k}\}$,  $1\leq k\leq \left[N/2\right]$ with two distinct spinors. Dicke states are characterized by two orthogonal spinors and belong to this family -  but they are not spin squeezed. On the otherhand, those constituted by two non-orthogonal spinors exhibit spin squeezing. 

\noindent{{\bf Keywords:}} Spin squeezing; Majorana geometric representation; Symmetric multiqubit states
\end{abstract} 
\pacs{03.65.Ud, 03.67.-a}
\maketitle 
\section{Introduction}
Spin squeezing in multiqubit states has been an intense 
area of research~\cite{wkz,ku,wn1,wn2,caves,yurke,ku2,gsapuri,puri,sor,ulamku,um,usha1,usha2,s1,s2,us1,us2,dyan,akru,gtoth,song,dsu} both for its theoretical importance and for its applicability in entanglement-enhanced sensing in quantum metrology. Spin squeezing implies pairwise entanglement~\cite{sor,ulamku,usha1,usha2,s1,s2, Vidal04, Vidal06} and has gained significance in the  quantification of metrologically useful entanglement in large number of ensembles of atomic spins.   

Kitegawa and Ueda~\cite{ku} proposed a definition of spin squeezing in terms of the uncertainty relation between collective angular momentum components of a spin $j=N/2$ state of an arbitrary $N$-qubit system.  A quantitative measure of spin squeezing, incorporating invariance under rotation of the frame of reference, is defined as~\cite{ku}, $\xi=2\left(\Delta{J}_\bot\right)_{\rm min}/\sqrt{N}$ and  a $N$-qubit state is spin squeezed if the {\emph {minimum}} value of the variance  $(\Delta{J_\bot})^2$ of the spin component $J_\bot$, in the direction perpendicular  to the mean spin direction $\langle \vec{J}\rangle$,  is smaller than the standard quantum limit $N/4$ of the coherent spin states~\cite{ku}.  More specifically,  
\be
\left(\Delta{J_\bot}\right)^2_{{\rm min}} \leq \frac{N}{4}\Longrightarrow\, \frac{2\left(\Delta{J}_\bot\right)_{\rm min}}{\sqrt{N}}\leq 1
\ee
and hence the  parameter  $\xi=\frac{2\left(\Delta{J}_\bot\right)_{{\rm min}}}{\sqrt{N}}<1$ for spin squeezed states~\cite{ku}. 
Here $\left(\Delta{J}_\bot\right)^2=\left\langle {J}_\bot^2\right\rangle - \left\langle {J}_\bot\right\rangle^2$
is the variance of the component  ${J}_\bot$ of the angular momentum, perpendicular to the mean spin-direction $\hat{n}_0~=~\frac{\langle\vec{J}\rangle}{\left|\langle\vec{J}\rangle\right|}~=~\frac{\left(\langle{\hat{J}}_x\rangle,\,\langle{\hat{J}}_y\rangle,\,\langle{\hat{J}}_z\rangle\right)}{\sqrt{\langle{\hat{J}}_x\rangle^2+\langle{\hat{J}}_y\rangle^2+\langle{\hat{J}}_z\rangle^2}}$, of the $N$-qubit system. 
The spin-squeezing parameter of a $N$-qubit symmetric state can be expressed in terms of the correlation matrix elements of its two-qubit reduced density matrix, expressed in the standard form~\cite{usha2}, 
\begin{eqnarray}
\label{st1}
\rho&=&\frac{1}{4}\, \left[I\otimes I+\sum_{i=x,y,z} \left(\sigma_{i}\otimes I + I\otimes \sigma_i\right) \, s_{i}  \right.\nonumber \\
&& \ \ \left. +\sum_{i,j=x,y,z} (\sigma_{i}\otimes \sigma_{j})\,t_{ij}\,\right]\, , 
\label{rho} 
\end{eqnarray}
where $I$ denotes the $2\times 2$ unit matrix,  $\sigma_i$ are the standard Pauli spin matrices and 
\begin{eqnarray}
\label{st2}
s_{i}&=&{\rm Tr}\,\left[\rho\, (\sigma_i\otimes I)\right]={\rm Tr}\,\left[\rho\, (I\otimes \sigma_i)\right] \nonumber \\ 
t_{ij}&=&{\rm Tr}\,\left[\rho\, (\sigma_{i}\otimes\sigma_{j})\right]={\rm Tr}\,\left[\rho\, (\sigma_{j}\otimes\sigma_{i})\right]=t_{ji}, 
\end{eqnarray} 
denote the qubit orientations $\vec{s}=(s_x,\,s_y,\,s_z)$ and  the real symmetric $3\times 3$ correlation matrix  $T=(t_{ij}), \ \ i,j=x,y,z$.  
The spin-squeezing parameter $ \xi$ can be expressed in the following simple form~\cite{usha2}: 
\begin{eqnarray}
\label{xitil1}
{\xi}&=&\left[1+(N-1)(\widetilde{\hat n}_{\bot}\,T\, 
{\hat n}_{\bot})_{\rm min}\right]^{1/2}.  
\end{eqnarray}
where ${\hat n}_\bot$ is a unit vector perpendicular to the mean spin direction ${\hat n}_0$ and  $\widetilde{\hat n}_{\bot}$ denotes its transpose.  
Choosing a suitable co-ordinate system with mutually orthogonal triads $\hat{n}_{1},\, \hat{n}_2,\, \hat{n}_0$ of basis vectors, such that the $Z$-axis is aligned along the unit vector $\hat n_0$ (mean spin direction), it may be seen that  the quadratic form $({\widetilde {\hat n}_{\bot}}{T}{\hat n}_{\bot})_{\rm min}$ is the {\em minimum eigenvalue}~\cite{usha2,dsu} of the $2\times 2$ block $T_\bot$  of the correlation matrix $T$ in the basis ${\hat n}_{1},\hat{n}_{2}$ orthogonal to the mean spin direction $\hat{n}_0$:
\begin{eqnarray}
\label{nmin}
&&({\widetilde{\hat n}_{\bot}}{T}{\hat n}_{\bot})_{\rm min}=\left.\frac{1}{2}\, \right[ 
\left( \widetilde{\hat n}_1\, T\,{\hat n}_{1}+\widetilde{\hat n}_2\, T\, 
{\hat n}_{2}\right)  \nonumber \\ 
&&  \left.\ \  - \sqrt{\left(\widetilde{\hat n}_1\,T\,{\hat n}_{1}-\widetilde{\hat n}_2\,T\,{\hat n}_{2}\right)^2 +  4\, \left(\,\widetilde{\hat n}_1 \,T \,{\hat n}_{2}\right)^2}\right].
\end{eqnarray}
Consequently, the spin-squeezing parameter $\xi$ can be expressed in an operationally simple form, on substituting Eq.~(\ref{nmin}) into  Eq.~(\ref{xitil1}). In other words, the Kitegawa-Ueda spin-squeezing parameter $\xi$ may be evaluated using  the   two-qubit reduced density matrix of {\em  any} random pair of qubits of a $N$-qubit symmetric system. 
 
In this paper we employ  Majorana geometric representation~\cite{majorana} of pure symmetric $N$-qubit states and explore spin squeezing in the family 
$\{{\cal D}_{N-k,k}\}$ of $N$-qubit states with two distinct Majorana spinors. Dicke states, the common eigenstates of collective operators ${\hat J}^2$, ${\hat J}_z$, are special states of the family $\{{\cal D}_{N-k,k}\}$,  ($1\leq k\leq \left[N/2\right]$) with the constituent  two distinct orthogonal spinors  $\vert 0 \rangle$, $\vert 1\rangle$. It is well known that Dicke states are entangled but are not spin squeezed~\cite{gtoth2}. We focus our attention to investigate spin squeezing behavior of $N$-qubit symmetric states constituted by two distinct {\em non-orthogonal} spinors belonging to the Majorana family $\{{\cal D}_{N-k,k}\}$, where one of the spinors occurs $k$ times and the other $(N-k)$ times. We evaluate the reduced two-qubit density matrix of $N$-qubit states belonging to different SLOCC classes~\cite{arus,bastin,solano} of the family $\{{\cal D}_{N-k,k}\}$ with $1\leq k\leq \left[N/2\right]\}$ and deduce the spin squeezing parameter for different values of $k$.  

This paper is organized as follows: Section~II gives an overview of the Majorana geometric representation of symmetric $N$-qubit states and provides a canonical structure for the family of states $\{{\cal D}_{N-k,k}\}$ with two distinct spinors. In Section III, we evaluate the 
two-qubit reduced density matrices and the spin-squeezing parameter of the $N$-qubit states belonging to $\{{\cal D}_{N-k,k}\}$, with  $1\leq k\leq \left[N/2\right]$. The variation of the spin squeezing parameter for the family of states $\{{\cal D}_{N-k,k}\}$, with different values of $k=1,2,\ldots ,[N/2]$ and $N$, is illustrated in Section~III. Section IV contains a brief summary.

 \section{Majorana representation of pure symmetric multiqubit states}
In the novel 1932 paper~\cite{majorana} Ettore Majorana proposed that a quantum system prepared in a  pure spin $j=\frac{N}{2}$ state can be represented as a permutation of the states of  $N$ constituent qubits  as follows:  
\begin{equation}
\label{Maj}
\vert \Psi_{\rm sym}\rangle={\cal N}\, \sum_{P}\, \hat{P}\, \{\vert \epsilon_1, \epsilon_2, 
\ldots  \epsilon_N \rangle\}, 
\end{equation} 
where 
\begin{equation}
\label{spinor}
\vert\epsilon_l\rangle= a_l  \vert 0\rangle + b_l e^{i\beta_l} \, \vert 1\rangle,\ \ l=1,2,\ldots, N, \ \ \ a_l^2+b_l^2=1
\end{equation}
denote the states of the qubits (spinors) constituting the symmetric $N$-qubit state $\vert \Psi_{\rm sym}\rangle$;
 $\hat{P}$ corresponds to the set of all $N!$ 
permutations and ${\cal N}$ corresponds to an overall normalization factor. Eq.~(\ref{Maj}) is referred  to as the {\em Majorana geometric representation} of a pure quantum state $\vert \Psi_{\rm sym}\rangle$ of spin $j=N/2$ or equivalently, that of permutationally symmetric $N$ qubits, expressed in terms of the constituent spinors $\vert \epsilon_l \rangle,\  l=1,2,\ldots N$. It may be  seen that when all the $N$ spinors $\vert\epsilon_l\rangle,\ \ l=1,2,\ldots, N$  are identical, the corresponding class $\{{\cal D}_{N}\}$ consists of separable states $\vert D_N\rangle=\vert \epsilon,\epsilon,\ldots \epsilon\rangle$.  The states  
\begin{eqnarray}
\label{n1n2}
\vert D_{N-k,k}\rangle&=&{\cal N}\,[\vert \underbrace{\epsilon_1, \epsilon_1,\ldots 
\epsilon_1}_{N-k}, \underbrace{\epsilon_2, \epsilon_2,\ldots \epsilon_2}_{k}\rangle+ \nonumber \\ 
& +&{\rm \, permutations\,}],  \ \ k=1,2, \ldots [N/2]
\end{eqnarray} 
constructed from two distinct spinors $\vert\epsilon_1\rangle,\ \vert\epsilon_2\rangle$ belong to the family $\{{\cal D}_{N-k,k}\}$ of $N$ qubits. It may be noted that the Dicke states $\left\vert \frac{N}{2}, \frac{N}{2}-k\right\rangle$, 
$k=1,2, \ldots [N/2]$ are the representative states of the family $\{{\cal D}_{N-k,k}\}$, with two orthogonal spinors $\vert\epsilon_1\rangle=\vert 0\rangle$, $\vert\epsilon_2\rangle=\vert 1\rangle$.   

An arbitrary  symmetric state  belonging to the family $\{ {\cal D}_{N-k,k}\}$ is given by~\cite{arus,bastin,solano},  
\begin{eqnarray*}
\vert D_{N-k, k}\rangle &=& {\cal N}\, \sum_{P}\, \hat{P}\,\{ \vert \underbrace{\epsilon_1, \epsilon_1,
\ldots , \epsilon_1}_{N-k};\ \underbrace{\epsilon_2, \epsilon_2,\ldots , \epsilon_2}_{k}\rangle\}
\end{eqnarray*} 
 and it can be reduced to a canonical form, characterized by only one real parameter~\cite{arus,arss}, with the help of identical local unitary transformations on individual  qubits. More specifically,   symmetric pure states $\vert D_{N-k, k}\rangle$ belonging to the family $\{{\cal D}_{N-k}\}$, are equivalent (under local unitary transformations) to the canonical state of the following form~\cite{arss} 
\begin{eqnarray}.
\label{nono}
\vert D_{N-k, k}\rangle &\equiv & \sum_{r=0}^k\, \beta^{(k)}_{r}\,\,  \left\vert\frac{N}{2},\frac{N}{2}-r \right\rangle, \\ 
\beta^{(k)}_{r}&=&{\cal N}\,\, 
\sqrt{\frac{N!(N-r)!}{r!}}\,\frac{a^{k-r}\, b^r}{(N-k)! (k-r)!}\,  \nonumber 
 \end{eqnarray}
where  $0\leq a, b=\sqrt{1-a^2}\leq 1$ are real parameters.  

In the next section we evaluate the two-qubit reduced density matrix of the state $\vert D_{N-k, k}\rangle$ for different values of  $k=1,2,\ldots $, for any  $N$  and deduce the spin-squeezing parameter $\xi$ (see Eq.~(\ref{xitil1})) corresponding to inequivalent classes $k=1,2,\ldots [N/2]$ of the family  $\{{\cal D}_{N-k,k}\}$. We establish that  the states $\vert D_{N-k,k}\rangle$ are spin-squeezed, except when $a=0$ and $a=1$.

\section{Spin squeezing in the family $\{ {\cal D}_{N-k,k}\}$ of $N$-qubit symmetric states with two distinct spinors}

In order to analyze spin squeezing in the different inequivalent SLOCC classes~\cite{arus,bastin,solano}, corresponding to $k=1,\,2,\,3,\,\ldots, [N/2]$, in the family $\{ {\cal D}_{N-k,k}\}$ of symmetric states, we first obtain the two-qubit reduced  density matrix $\rho^{(k)}$ corresponding to any random pair of qubits in the state $\vert D_{N-k,\,k}\rangle\in \{ {\cal D}_{N-k,k}\} $. We have  
\begin{widetext}
\begin{eqnarray}
\label{formal_rhok}
\rho^{(k)}&=&\mbox{Tr}_{N-2}\left(\vert D_{N-k,\,k}\rangle\langle D_{N-k,\,k} \vert\right) \nonumber \\ 
&=&\mbox{Tr}_{N-2}\,\left\{  \sum_{r,r'=0}^k\, \beta^{(k)}_r\, \beta^{(k)}_{r'} \sum_{m_2,m_2'}\,\left[ c_{m_2}^{(r)}\, 
c_{m'_2}^{(r')}\,
\left\vert\frac{N}{2}-1,\frac{N}{2}-r-m_2 \right\rangle 
\left\langle \frac{N}{2}-1,\frac{N}{2}-r'-m_2'   \right\vert \otimes \vert 1, m_2\rangle \langle 1, m_2'\vert \right]  \right\} \nonumber \\ 
&=& \sum_{m_2,m_2'=1,0,-1}\, \rho^{(k)}_{m_2,m_2'}\,  \vert 1, m_2\rangle \langle 1, m_2'\vert,  
\end{eqnarray}
where 
\begin{eqnarray}
\label{rhok}
\rho^{(k)}_{m_2,m_2'}=\sum_{r,r'=0}^k\, \beta^{(k)}_r\, \beta^{(k)}_{r'}\, c_{m_2}^{(r)}\,c_{m'_2}^{(r')}\,\sum_{m_1=(-N/2)+1}^{(N/2)-1}\, \left\langle \frac{N}{2}-1, m_1    \right\vert\left.\frac{N}{2}-1,\frac{N}{2}-r-m_2 \right\rangle 
\left\langle \frac{N}{2}-1,\frac{N}{2}-r'-m_2'   \right\vert  \left.\frac{N}{2}-1, m_1\right\rangle \nonumber \\
\end{eqnarray}
\end{widetext}
The associated Clebsch-Gordan coefficients 
$c^{(r)}_{m_2}~=~C\left(\frac{N}{2}-1,\, 1,\, \frac{N}{2};m-m_2,\, m_2, m \right)$, $m~=~\frac{N}{2}-r$, $m_2=1,\,0,\,-1$ are given explicitly by~\cite{Var}
\begin{eqnarray}
\label{cg_explicit}
c^{(r)}_{1}&=&\sqrt{\frac{(N-r)(N-r-1)}{N(N-1)}},\, c^{(r)}_{-1}=\sqrt{\frac{r\, (r-1)}{N(N-1)}}   \nonumber \\
&&\ \  \ \ c^{(r)}_{0}=\sqrt{\frac{2r\, (N-r)}{N(N-1)}}	  
\end{eqnarray}
By expressing $\rho^{(k)}$ in the standard two-qubit basis $\{\vert 0_A,0_B \rangle, \vert 0_A,1_B \rangle, \vert 1_A,0_B \rangle, \vert 1_A,1_B \rangle\}$, (using the relations between angular momentum basis $\vert 1,m_2=\pm 1, 0\rangle$ and the local qubit basis i.e.,  
$\vert 1,1\rangle=\vert 0_A,0_B\rangle$,\  $\vert 1,0\rangle=(\vert 0_A,1_B\rangle+\vert 1_A,0_B\rangle)/\sqrt{2}$,\ $\vert 1,-1\rangle=
\vert 1_A,1_B\rangle$), one obtains the following simplified form~\cite{s2} for the symmetric two-qubit reduced density matrix:     
\begin{eqnarray}
\label{rhok_matrix}
\rho^{(k)}&=&\ba{cccc} A^{(k)} \ \ & B^{(k)} \ \ & B^{(k)}\ \  & C^{(k)} \ \  \\ B^{(k)} \ \  & D^{(k)}\ \  & D^{(k)}\ \  & E^{(k)} \ \   \\ B^{(k)} \ \  & D^{(k)} \ \  & D^{(k)} \ \ & E^{(k)} \ \  \\ C^{(k)} \ \  & E^{(k)} \ \  & E^{(k)} \ \  & F^{(k)} \ \  \ea, 
\end{eqnarray}
where $A^{(k)},\, B^{(k)},\, C^{(k)},\, D^{(k)},\, E^{(k)}$ and $F^{(k)}$ are real.
   
Now we proceed to discuss spin squeezing in detail in the illustrative cases $k=1,2$ in the family of states $\{{\cal D}_{N-k,\,k}\}$.   

\subsection{Spin squeezing in the class of states $\{{\cal D}_{N-1,\,1}\}$}

The reduced two-qubit density matrix $\rho^{(1)}$ drawn from the $N$-qubit pure states of the family  $\{{\cal D}_{N-1,\,1}\}$ (see Eq.~(\ref{formal_rhok})) has the following explicit structure: 
\begin{eqnarray}
\label{rho1}
&&\rho^{(1)}=\mbox{Tr}_{N-2}\left(\vert D_{N-1,\,1}\rangle\langle D_{N-1,\,1}\right)\nonumber \\
&&\ = \left( \left(\beta^{(1)}_0\right)^2+ \left(\beta^{(1)}_1\,  c^{(1)}_1\right)^2   \right)\vert 1,\,1\rangle\langle 1,\,1 \vert \nonumber \\ 
&& \ \ \  + \left(\beta^{(1)}_1\, c^{(1)}_0\right)^2  
\vert 1,\,0\rangle\langle 1,\,0 \vert  +\beta^{(1)}_0 \beta^{(1)}_1\, c^{(1)}_0 \vert 1,\,1\rangle\langle 1,\,0 \vert \nonumber \\ 
&&\ \ \  +\beta^{(1)}_0 \beta^{(1)}_1\, c^{(1)}_0 \vert 1,\,0\rangle\langle 1,\,1 \vert 
\end{eqnarray}
Here  (see Eq. (\ref{nono})) we have $\beta^{(1)}_0={\cal N}N\, a $, $\beta^{(1)}_1={\cal N}\, \sqrt{N(1- a^2)}$ with ${\cal N}=\frac{1}{\sqrt{N^2\,a^2+N(1-a^2)}}$ and  the associated non-zero Clebsch-Gordan coefficients (see Eq. (\ref{cg_explicit})) are given by 
 \begin{equation}
 \label{cg_explicit1}
 c^{(1)}_1=\sqrt{\frac{N-2}{N}},\  c^{(1)}_0=\sqrt{\frac{2}{N}}.
\end{equation}
Furthermore, in the standard two-qubit basis $\{\vert 0_A,0_B \rangle, \vert 0_A,1_B \rangle, \vert 1_A,0_B \rangle, \vert 1_A,1_B \rangle\}$, we obtain   
\begin{eqnarray}
\label{rho1_matrix}
\rho^{(1)}&=&\ba{cccc} A^{(1)} \ \ & B^{(1)} \ \ & B^{(1)}\ \  & 0 \ \  \\ B^{(1)} \ \  & D^{(1)}\ \  & D^{(1)}\ \  & 0 \ \   \\ B^{(1)} \ \  
& D^{(1)} \ \  & D^{(1)} \ \ & 0 \ \  \\ 0 \ \  & 0 \ \  & 0 \ \  & 0 \ \  \ea \nonumber
\end{eqnarray}
where 
\begin{eqnarray}
A^{(1)}&=&\frac{N^2a^2+(N-2)(1-a^2)}{N^2\,a^2+N(1-a^2)},\ B^{(1)}=\frac{a\sqrt{1-a^2}}{1+a^2(N-1)},\   \nonumber \\
 D^{(1)}&=& \frac{1-a^2}{N^2\,a^2+N(1-a^2)},\ 
 \end{eqnarray}
We obtain the qubit orientations (see (Eq. \ref{st2}))  
\begin{eqnarray*}
s_x= 2\,B^{(1)}, \ s_y=  0, \ s_z= A^{(1)}  
\end{eqnarray*}
using which we find an orthogonal triad of basis vectors
\begin{eqnarray*}
\hat{n}_0&=&\left(\frac{s_x}{\sqrt{s_x^2+s_z^2}},0,\frac{s_z}{\sqrt{s_x^2+s_z^2}}
\right), 
\\
\hat{n}_1&=& (0,\,1,\,0), \nonumber \\
\hat{n}_2&=&\left(-\frac{s_z}{\sqrt{s_x^2+s_z^2}}, 0, \frac{s_x}{\sqrt{s_x^2+s_z^2}} \right)
\end{eqnarray*}
with $\hat{n}_0$ denoting the mean spin direction.  On simplifying, we obtain (see Eq. (\ref{nmin})) $\widetilde{\hat n}_{1}\, T\,{\hat n}_{2}=\widetilde{\hat n}_{2}\, T\,{\hat n}_{1}=0$ and 
$({\widetilde{\hat n}_{\bot}}{T}{\hat n}_{\bot})_{{\rm min}}= \widetilde{\hat n}_{2}\, T\,{\hat n}_{2}$. 
Thus the corresponding spin-squeezing parameter takes the form~\cite{note} (see (Eq. \ref{xitil1}))  
\begin{eqnarray}
\label{xin-11}
\xi&=&\sqrt{1+(N-1)\,   ({\widetilde{\hat n}}_{2}\, T{\hat n}_{2})\, 
}
\end{eqnarray}
where 
\be
\widetilde{\hat n}_{2}\, T\,{\hat n}_{2}  
=\frac{2\left[\left(A^{(1)}\right)^2\,D^{(1)}-2\,\left(B^{(1)}\right)^2 \right]}{4\,\left(B^{(1)}\right)^2+\left(A^{(1)}\right)^2}.     
\ee
In Fig.~1 we have plotted  $\xi$, for the states in the family $\{{\cal D}_{N-1,1}\}$, as a function of the parameter $a$ and number of qubits $N$. 

\begin{figure}[h]
	\label{d11}
	\begin{center}
		\includegraphics*[width=3in,keepaspectratio]{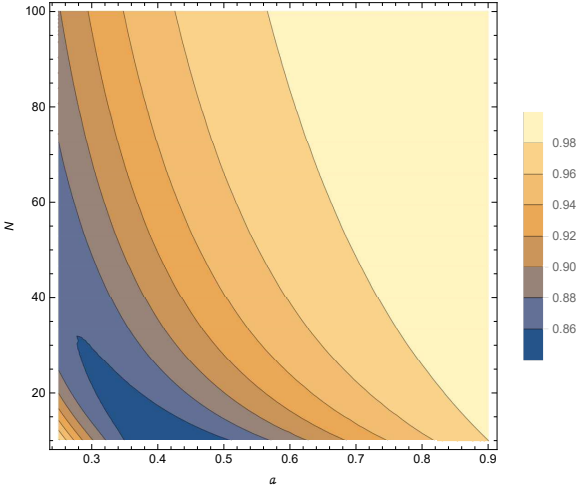}
		\caption{Spin-squeezing parameter $\xi$ as a function of the parameter $a$ and number of qubits $N$ in the class  $\{{\cal D}_{N-1,1}\}$ of $N$-qubit pure symmetric states with two distinct spinors.} 
	\end{center}
\end{figure}  


\subsection{Spin squeezing in the class $\{{\cal D}_{N-2,\,2}\}$} 
\label{ss3B}
When $k=2$, the state $\vert D_{N-k, k}\rangle$ has the form (see Eq. (\ref{nono}))
\begin{eqnarray}
\vert D_{N-2, 2}\rangle 
&=& \beta^{(2)}_0\,\left\vert\frac{N}{2},\frac{N}{2} \right\rangle+\beta^{(2)}_1\,\left\vert\frac{N}{2},\frac{N}{2}-1 \right\rangle\nonumber \\ 
&& \ +\beta^{(2)}_2\,\left\vert\frac{N}{2},\frac{N}{2}-2 \right\rangle
\end{eqnarray}
where
\begin{eqnarray}
\beta^{(2)}_0&=&{\cal N}\frac{N(N-1)}{2} a^2 \nonumber \\ 
\beta^{(2)}_1&=&{\cal N}\sqrt{N} (N-1) a \sqrt{1-a^2} \\
\beta^{(2)}_2&=&{\cal N}\sqrt{\frac{N(N-1)}{2}} (1-a^2).  \nonumber 
\end{eqnarray}
and ${\cal N}$, the normalization factor, satisfies the relation $(\beta^{(2)}_0)^2+(\beta^{(2)}_1)^2+(\beta^{(2)}_2)^2=1$.
Following the procedure outlined in subsection IIIA, we evaluate the two-qubit reduced density matrix $\rho^{(2)}=\mbox{Tr}_{N-2}\left(\vert D_{N-2,\,2}\rangle\langle D_{N-2,\,2} \vert\right)$ and express it in the standard two-qubit basis $\{\vert 0_A,0_B \rangle, \vert 0_A,1_B \rangle, \vert 1_A,0_B \rangle, \vert 1_A,1_B \rangle\}$: 
\begin{eqnarray}
\label{rho2_matrix}
\rho^{(2)}&=&\ba{cccc} A^{(2)} \ \ & B^{(2)} \ \ & B^{(2)}\ \  & C^{(2)} \ \  \\ B^{(2)} \ \  & D^{(2)}\ \  & D^{(2)}\ \  & E^{(2)} \ \   \\ B^{(2)} \ \  & D^{(2)} \ \  & D^{(2)} \ \ & E^{(2)} \ \  \\ C^{(2)} \ \  & E^{(2)} \ \  & E^{(2)} \ \  & F^{(2)} \ \  \ea \nonumber
\end{eqnarray}
where 
\begin{eqnarray*} 
A^{(2)}&=&\left(\beta^{(2)}_0\right)^2+\left(\beta^{(2)}_1\,  c^{(1)}_1\right)^2+ \left(\beta^{(2)}_2\, c^{(2)}_1\right)^2,\nonumber \\ 
B^{(2)}&=&\frac{\beta^{(2)}_0 \beta^{(2)}_1 c^{(1)}_0+\beta^{(2)}_1 \beta^{(2)}_2 c^{(1)}_1 c^{(2)}_0}{\sqrt{2}}, \nonumber \\ 
C^{(2)}&=&\beta^{(2)}_0 \beta^{(2)}_2 c^{(2)}_{-1}, \nonumber \\ 
D^{(2)}&=& \frac{\left(\beta^{(2)}_1  \, c^{(1)}_0\right)^2+\left(\beta^{(2)}_2\, c^{(2)}_0\right)^2 }{2}, \nonumber \\ 
E^{(2)}&=& \frac{\beta^{(2)}_1 \beta^{(2)}_2 c^{(1)}_{0} c^{(2)}_{-1}}{\sqrt{2}}, \nonumber \\ 
F^{(2)}&=& \left(\beta^{(2)}_2\, c^{(2)}_{-1}\right)^2, 
\end{eqnarray*}
and the associated  non-zero Clebsch-Gordan coefficients (see Eq. (\ref{cg_explicit})) are given in Eq. (\ref{cg_explicit1}) and  
\begin{eqnarray*} 
c^{(2)}_1&=&
\sqrt{\frac{(N-3)(N-2)}{N(N-1)}},\  
c^{(2)}_0=2\, \sqrt{\frac{N-2}{N(N-1)}} \\
c^{(2)}_{-1}&=&\sqrt{\frac{2}{N(N-1)}}. \nonumber	
	\end{eqnarray*}
Substituting for $\beta^{(2)}_i$, $i=0,\,1,\,2$ and the Clebsch-Gordan coefficients, we obtain the density matrix $\rho^{(2)}$  in terms of 
 the number $N$ of qubits, and the real  parameter $a$.  We then evaluate the spin-squeezing parameter $\xi$  following the same procedure followed in subsection~IIIA while discussing the class $\{{\cal D}_{N-1,1}\}$. We identify~\cite{note} that the mean spin direction $\hat{n}_0$  lies in the XZ-plane and the element of correlation matrix $\widetilde{\hat{n}}_1\, T \, \hat{n}_{2}=0.$ This facilitates the evaluation of the spin squeezing parameter to be
 $\xi=\sqrt{1+(N-1)\,  ({\widetilde{\hat n}}_{2}\, T\,{\hat n}_{2}),\, 
 }.$ 
 We have plotted  $\xi$ as a function of  the parameter $a$
for different values of $N$ in Fig~2. 
\begin{figure}[h]
\label{d2}
\begin{center}
\includegraphics*[width=3in,keepaspectratio]{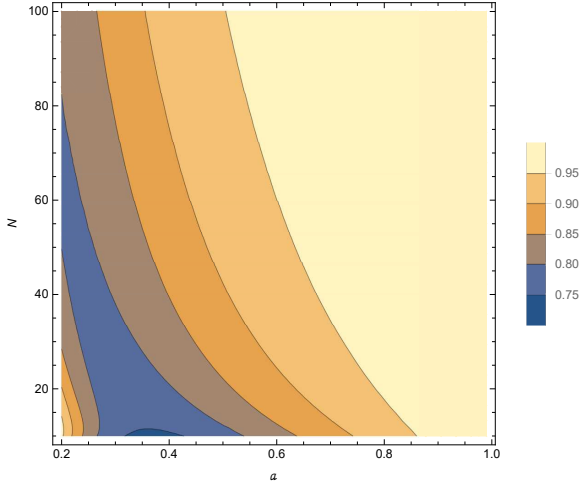}
\caption{Spin-squeezing parameter $\xi$ as a function of the parameter $a$ and number of qubits $N$ in the class  $\{{\cal D}_{N-2,2}\}$ of pure symmetric states.} 
\end{center}
\end{figure} 

In general, for any arbitrary $k$, we evaluate the two-qubit density matrix  $\rho^{(k)}=\mbox{Tr}_{N-2}\left(\vert D_{N-k,\,k}\rangle\langle D_{N-k,\,k} \vert\right)$,  in the standard two-qubit basis, in the form given in Eq. (\ref{rhok_matrix}) with elements $A^{(k)},\ B^{(k)},\ C^{(k)},\ D^{(k)},\  E^{(k)},\ F^{(k)}$ given by,  
\begin{eqnarray*}
 A^{(k)}&=&\sum_{r=0}^k \left(\beta_r^{(k)}\,c_1^{(r)}\right)^2,\\    B^{(k)}&=&\frac{1}{\sqrt{2}}\sum_{r=0}^{k-1}\,\beta_{r}^{(k)}\, \beta_{r+1}^{(k)}\, {c^{(r)}_{1}}\, {c^{(r+1)}_{0}},  \\ 
C^{(k)}&=&\sum_{r=0}^{k-2}\,\beta_{r}^{(k)}\, \beta_{r+2}^{(k)}\, {c^{(r)}_{1}}\, {c^{(r+2)}_{-1}}, \nonumber \\ 
D^{(k)}&=&\frac{1}{2}\sum_{r=0}^{k-1}\,|\beta_{r+1}^{(k)}|^{2}\, {c^{(r+1)}_{0}}^{2},\ 
\nonumber \\
 E^{(k)}&=&\frac{1}{\sqrt{2}}\sum_{r=0}^{k-2}\,\beta_{r+1}^{(k)}\, \beta_{r+2}^{(k)}\, {c^{(r+1)}_{0}}\, {c^{(r+2)}_{-1}},\ \  \nonumber \\  
F^{(k)}&=&\sum_{r=0}^{k-2}\,\left(\beta_{r+2}^{(k)}\, c^{(r+2)}_{-1}\right)^{2}.
\end{eqnarray*}
The mean spin direction $\hat{n}_0$ of the qubits in $\vert D_{N-k, k}\rangle$ lies in the XZ-plane and the spin-squeezing parameter $\xi$ for any arbitrary state $\vert D_{N-k,k}\rangle$ belonging to the family $\{ {\cal D}_{N-k,k}\}$ can be readily  evaluated~\cite{note} using 
$\xi~=~\sqrt{1+(N-1)\,(\widetilde{\hat{n}}_2\, T\, \hat{n}_2})\,$ where 
\begin{eqnarray*}
\widetilde{\hat n}_{2}\, T\,{\hat n}_{2}  
&=& \frac{2\left(A^{(k)}-F^{(k)}\right)^2\,\left(C^{(k)}+D^{(k)}\right)
}
{\left[4\,\left(B^{(k)}+E^{(k)}\right)^2+\left(A^{(k)}-F^{(k)}\right)^2\right]} \\	
&&	\ \ \ 	+\frac{
	4\,\left(B^{(k)}+E^{(k)}\right)^2\left(1-4D^{(k)}\right)}
	{\left[4\,\left(B^{(k)}+E^{(k)}\right)^2+\left(A^{(k)}-F^{(k)}\right)^2\right]}\\ 
&&\ \ 	-		\frac{8\, \left(A^{(k)}-F^{(k)}\right)\left(\left(B^{(k)}\right)^2-\left(E^{(k)}\right)^2
			\right)}
	{\left[4\,\left(B^{(k)}+E^{(k)}\right)^2+\left(A^{(k)}-F^{(k)}\right)^2\right]}.
\end{eqnarray*} 
In Fig.~3 we have illustrated the variation of the spin-squeezing parameter $\xi$ in the $N$-qubit pure symmetric state $\vert D_{N-k,k}\rangle$ with different $k$ and $N$. 
\begin{figure}[h]
\label{com15}
\begin{center}
\includegraphics*[width=3.5in,keepaspectratio]{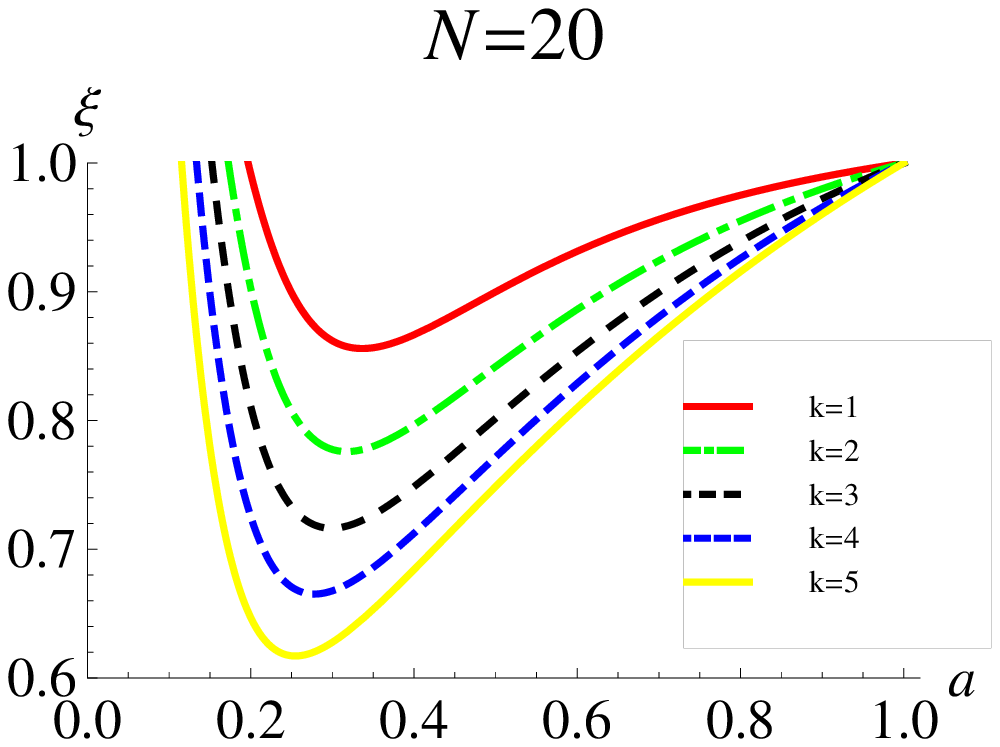} 
\includegraphics*[width=3.5in,keepaspectratio]{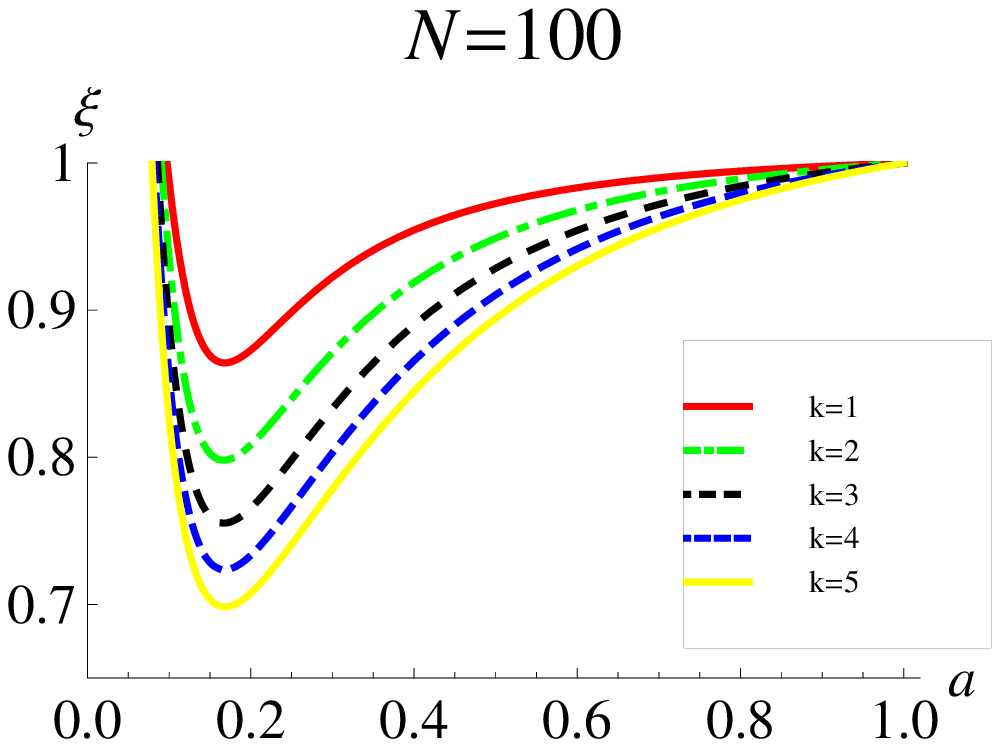}
\caption{Spin squeezing parameter $\xi$ of states $\vert D_{N-k, k}\rangle$ (see (Eq. \ref{nono})) for different values of $k=1,2,3,4,5$ and for number of qubits $N=20$ and   $N=100$. } 
\end{center}
\end{figure} 
It can be readily seen from Fig.~3 that for a fixed $N$, the spin-squeezing parameter for the state $\vert D_{N-k,k}\rangle$  reduces with the increase in $k$.  

\section{Concluding remarks} 
In this article we have explored spin squeezing in symmetric multiqubit pure states belonging to the family of two distinct Majorana spinors. We exclusively make use of the fact that spin squeezing is a reflection of pairwise entanglement and evaluation of the  spin squeezing parameter requires the knowledge of two-qubit reduced density matrix of the $N$-qubit symmetric state~\cite{usha2}. We have used the canonical form of pure symmetric states $\vert D_{N-k,k}\rangle$ of $N$ qubits with two distinct spinors which are characterized by a single real parameter $a$ (see Eq. (\ref{nono})) and divided the system into two parts containing $N-2$ and $2$ qubits respectively. By tracing out the $N-2$ qubits we obtain the density matrix corresponding to any two qubits of the $N$-qubit symmetric state $\vert D_{N-k,k}\rangle$. The correlation matrix elements expressed in the basis perpendicular to the mean spin direction leads  us to evaluate the spin-squeezing parameter. The variation of spin squeezing with respect to the  real parameter $0< a<1$ characterizing the state $\vert D_{N-k,k}\rangle$ is graphically illustrated for SLOCC inequivalent family of states 
$\{ {\cal{D}}_{N-k,k}\}$, with different values of $k=1,2,\ldots$. While the Dicke states 
$\left\vert \frac{N}{2}, \frac{N}{2}-k\right\rangle$ constituted by  two orthogoanl spinors  are not spin squeezed, our work reveals that their generalizations viz.,  $N$-qubit symmetric  states, consisting of two non-orthogonal spinors, exhibit spin squeezing. Investigations on the metrological relevance of the $N$ qubit states belonging to the family of two distinct Majorana spinors is under progress and it will be presented in a separate communication~\cite{RA}.   

\section*{Acknowledgements} 
KSA would like to thank the University Grants Commission for providing a BSR-RFSMS fellowship during the present work. ARU acknowledges the support of UGC MRP (Ref. MRP-MAJOR-PHYS-2013-29318). 
We thank Professor A. K. Rajagopal for insightful discussions.


\begin{thebibliography}{0}
\bibitem{note} For states belonging to the class $\{ {\cal D}_{N-k,k}\}$ with different values of $k=2,3,\ldots $ we find that (i) the mean spin direction (denoted by the unit vector $\hat{n}_0$) lies in the XZ plane and  (ii) the $2\times 2$ block $T_\bot$ of correlation matrix $T$, which is expressed in the basis $({\hat n}_1, \hat{n}_2)$ orthogonal to the mean spin direction $\hat{n}_0$, is diagonal i.e.,  ${\widetilde{\hat n}_{1}}\, T \, {\hat n}_{2}=0$. It is seen that 
 ${\rm min}\left(\widetilde{\hat n}_{1}\, T\,
{\hat n}_{1},\, \widetilde{\hat n}_{2}\,T\,{\hat n}_{2}\right)=\widetilde{\hat n}_{2}\,T\,{\hat n}_{2}$. Thus the spin squeezing parameter  (see Eqs. (\ref{xitil1})   and (\ref{nmin})) takes the form $\xi=\sqrt{1+(N-1)\, \widetilde{\hat n}_{2}\,T\,{\hat n}_{2}}$  for any $k$ and $N$.
\end{thebibliography}
\end{document}